\documentclass[pre,aps,twocolumn, superscriptaddress]{revtex4}
\usepackage{amsmath}
\usepackage{epsfig}
\usepackage{amssymb}
\usepackage{color}

\begin{document}

\title{The probability distribution of Brownian motion in 
  periodic potentials}

\author{Matan Sivan} 
\affiliation{Department of Biomedical
Engineering, Ben-Gurion University of the Negev, Be'er Sheva 85105,
Israel} 

\author{Oded Farago} 
\affiliation{Department of Biomedical
Engineering, Ben-Gurion University of the Negev, Be'er Sheva 85105,
Israel} 
\affiliation{Ilse Katz Institute for Nanoscale Science and
Technology, Ben-Gurion University of the Negev, Be'er Sheva 85105,
Israel}

\begin{abstract}

We calculate the probability distribution function (PDF) of an
overdamped Brownian particle moving in a periodic potential energy
landscape $U(x)$. The PDF is found by solving the corresponding
Smoluchowski diffusion equation. We derive the solution for any
periodic even function $U(x)$, and demonstrate that it is
asymptotically (at large time $t$) correct up to terms decaying faster
than $\sim t^{-3/2}$. As part of the derivation, we also recover the
Lifson-Jackson formula for the effective diffusion coefficient of the
dynamics. The derived solution exhibits agreement with Langevin
dynamics simulations when (i) the periodic length is much larger than
the ballistic length of the dynamics, and (ii) when the potential
barrier $\Delta U=\max(U(x))-\min(U(x))$ is not much larger than the
thermal energy $k_BT$.

\end{abstract} 

\maketitle

\section{Introduction}

Brownian motion in a periodic potential constitutes one of the
fundamental problems of particle transport with numerous applications
in various fields of science and technology. Many classical examples
of diffusion in periodic systems are found in the area of condensed
matter physics, including diffusion of atoms in and on the surface of
lattices~\cite{weiner74,ala02}, and fluctuations of Josephson
supercurrent through a tunneling junction~\cite{ambegaokar69}. In many
situations, e.g., the cases of superionic conductors~\cite{fulde75}
and rotating dipoles in external fields~\cite{reguera00}, a constant
force which biases the stochastic dynamics in a given direction is
also present. In such scenarios, often referred to as diffusion in a
{\em tilted}\/ periodic potential different types of dynamics are
observed in the overdamped (high friction) regime depending on whether
the total potential energy (the sum of periodic and linear potentials)
has minima or not~\cite{reiman02}. In the former case, the particle
moves from one minimum to another and the solution is termed
``locked''; in the latter case, the particle moves down the corrugated
potential gradient and the solution in termed
``running''~\cite{risken}.

The problem of diffusion in periodic systems is also relevant to the
study of thermal
ratchets~\cite{reimannreview,hanggi09,cubero16}. Thermal ratchets
employ a time-varying spatially-asymmetric periodic potential that
drives isothermal systems out of equilibrium and allows for the
rectification of the thermal noise in the form of a directed
probability (particle) flux. Thermal ratchets attracted much renewed
interest in the 1990s' as possible models for motor
proteins~\cite{bier97}. Advances in various experimental techniques,
most notably in optical trapping (``tweezers'') devices, have led to
novel experimental setups where some of the new theoretical concepts
were tested~\cite{koss03}. Another closely-related problem attracting
considerable recent attention is diffusion in corrugated channels
\cite{zwanzig92}. Understanding Brownian motion in confined geometries
is important for the study of transport of materials in, e.g.,
zeolites~\cite{schuring02} and microfluidic
channels~\cite{weigl99}. Such dynamics can be studied by considering a
one-dimensional Fick-Jacobs diffusion equation of a particle moving in
the presence of an effective potential of mean force arising from the
variations in the cross-sectional area of the
channel~\cite{reguera01,kalinay06}.

Here, we consider one-dimensional diffusion in the high friction
regime. The probability distribution function (PDF), $P(x,t)$, of
finding the particle at coordinate $x$ at time $t$ can be found by
solving the Smoluchowski equation~\cite{smoluchpaper}
\begin{equation}
  \frac{\partial P(x,t)}{\partial t}=D\frac{\partial}{\partial
    x}\left\{e^{-\beta U(x)}\frac{\partial}{\partial x}
    \left[e^{\beta U(x)}P(x,t)\right]\right\},
  \label{eq:smoluch}
\end{equation}
where $D$ is the diffusion coefficient of the medium, $U(x)$ is the
potential energy function, and $\beta=1/k_BT$ where $T$ is the
temperature and $k_B$ is Boltzmann's constant. Throughout the paper, we
assume that $U(x)$ is an even periodic function with period $\lambda$
and consider a particle initially located at the origin, i.e.,
$P(x,t=0)=\delta(x)$, where $\delta$ is the Dirac delta-function. From
symmetry considerations, the mean displacement of the particle
vanishes identically, $\langle x\rangle=0$. The mean-squared
displacement does not vanish but rather exhibits, at asymptotically
large times, a linear growth with time characterizing the
diffusive nature of the dynamics. However, the effective diffusion
coefficient defined by
\begin{equation}
  D^*=\lim_{t\to\infty}\frac{\left\langle x^2\right\rangle}{2t},
  \label{eq:effectived}
\end{equation}
is {\em not}\/ equal to the medium diffusion coefficient, but is given
by the Lifson-Jackson (LJ) formula~\cite{lfpaper}
\begin{equation}
  D^*=\frac{D}{\left\langle e^{-\beta U(x)}\right\rangle
    \left\langle e^{\beta U(x)}\right\rangle},
  \label{eq:lifsonjackson}
\end{equation}
where $\langle \cdots \rangle$ denotes an average over a unit cell:
$\langle c \rangle=(\lambda)^{-1}\int_0^{\lambda} c(x)dx$. This
formula has been derived by several authors using somewhat different
approaches~\cite{festa78,gunther79,weaver79}, and it has been proved
that $D^*\leq D$~\cite{festa78}. Physically, the fact that $D^*\leq D$
is directly related to the tendency of the Brownian particle to get
trapped, for some duration, in the minima of the periodic potential
before moving to the adjacent cell.

We note here that the LJ formula is valid only when the potential
barrier, $\Delta U=U_{\rm max}-U_{\rm min}$ is not too high. In the
high barrier limit, $\beta \Delta U\gg 1$, the particle only rarely
escapes the vicinity of a potential minimum, and it advances to a
neighboring cell with a characteristic time that scales as $\tau\sim
\exp(\beta \Delta U)$. In this limit, the effective diffusion
coefficient is expected to follow an Arrhenius-Kramer
behavior~\cite{ferrando92}: $D^*\sim De^{-\beta \Delta U}$.

Despite the extensive theoretical literature on the problem of
diffusion in periodic potentials, the general solution of
Eq.~(\ref{eq:smoluch}) subject to the delta-function initial condition
is not known. Here, we take a major step forward and derive an
asymptotic (at large times) solution for the Smoluchowski equation for
any even periodic potential function $U(x)$. We use Langevin dynamics
simulations of a case-study to demonstrate excellent agreement between
our analytical solution and the computed PDF. As part of our
derivation, we independently arrive at the LJ formula for the
effective diffusion coefficient.

\section{Preliminary considerations}

Although we focus here on diffusive (overdamped) dynamics, our
computational results are actually based on inertial (underdamped)
Langevin dynamics simulations. The Langevin equation is given
by~\cite{langevinpaper}
\begin{equation}
  m\frac{dv}{dt}=-\alpha v+\xi(t)+f(x),
  \label{eq:langevin}
\end{equation}
where $m$ and $v$ denote, respectively, the mass and velocity of the
particle, $f=-dU(x)/dx$ is the deterministic force acting on the
particle, $\alpha> 0$ is the medium's friction coefficient, and
$\xi(t)$ is a thermal white nosie with zero average $\langle
\xi(t)\rangle=0$ and delta function auto-correlation
$\langle\xi(t)\xi(t^{\prime})\rangle=2k_BT\alpha\delta(t-t^{\prime})$~\cite{risken}. Langevin
dynamics is inertial at time scales much smaller than the ballistic
time $\tau_b\sim m/\alpha$, during which the particle moves a
characteristic distance $l_b\sim v_{\rm th}\tau_b$, where $v_{\rm
  th}=\sqrt{k_BT/m}$ is the thermal velocity of the particle. The
dynamics becomes diffusive at length scales much larger than
$l_b$. From Smoluchowski equation (\ref{eq:smoluch}) with
$D=k_BT/\alpha$ (Einstein's relation), one can derive the particle's
PDF in the overdamped limit of Langevin's equation, i.e., when
$l_b\rightarrow 0$. This description, however, is valid only if the
spatial variations of the deterministic force on length scales of the
order of $l_b$ are much smaller than the characteristic friction
force, i.e., for $(m/\alpha^2)|df/dx|\ll 1$~\cite{wilemski76}. In the
case of a periodic force $|df/dx|\sim \Delta U/\lambda^2$, implying
that Eq.~(\ref{eq:smoluch}) may not be valid when the periodic length
of the potential becomes comparable to the ballistic length, or in the
case when $\Delta U\gg k_BT$. We will later see that, indeed, in these
limits, the LJ formula derived from Eq.~(\ref{eq:smoluch}) fails to
depict correctly the effective diffusion coefficient $D^*$.

We notice that if $U(x)$ is periodic then the Boltzmann's weight
$\exp[-\beta U(x)]$ is also a periodic function.  We thus define the
periodic function $\eta(x)$
\begin{equation}
 1+\epsilon\eta(x)=\frac{e^{-\beta U(x)}}{\left \langle e^{-\beta
     U(x)}\right\rangle},
 \label{eq:eta}
\end{equation}
with the variable
\begin{equation}
  \epsilon=1-\frac{e^{-\beta U_{max}}}{\langle e^{-\beta U(x)}\rangle}
  \label{eq:epsilon}
\end{equation}
satisfying $0\leq\epsilon<1$. The function $\eta(x)$ has the following
properties: (i) $\langle\eta(x)\rangle=0$, and (ii) $\min [\eta(x)]
$=-1.  With the definition of $\epsilon$ and $\eta(x)$, the
Smoluchowski equation (\ref{eq:smoluch}) takes the form
\begin{equation}
  \frac{\partial P(x,t)}{\partial t}=D\frac{\partial}{\partial
    x}\left\{\left[1+\epsilon\eta(x)\right]\frac{\partial}{\partial x}
    \frac{P(x,t)}{\left[1+\epsilon\eta(x)\right]}\right\},
  \label{eq:smoluch2}
\end{equation}
and LJ formula (\ref{eq:lifsonjackson}) reads
\begin{equation}
  D^*=\frac{D}{ \left\langle
    \left[1+\epsilon\eta(x)\right]^{-1}\right\rangle}.
  \label{eq:lifsonjackson2}
\end{equation}
In the high barrier limit $\epsilon\rightarrow 1$, we expect the
Arrhenius-Kramer law which takes the form
\begin{equation}
  D^*\sim D(1-\epsilon).
  \label{eq:arrhen2}
\end{equation}

\begin{figure}[htb]
\centering\includegraphics[width=0.375\textwidth]{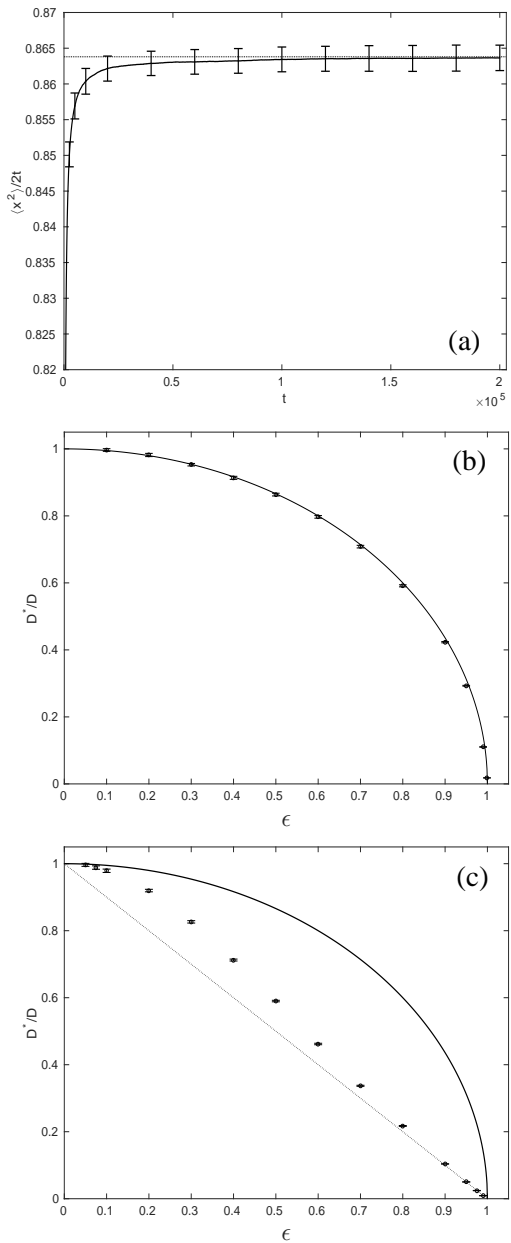} 
\caption{(a) The ratio between the mean squared displacement and twice
  the time, {\protect $D^*=\left\langle x^2\right\rangle/2t$}, as a
  function of $t$ for $\epsilon=0.5$ and $\lambda=50$. At large times
  this ratio converges to $D^*\simeq 0.863$ that matches the value for
  the effective diffusion coefficient $D^*$ predicted by the LJ
  formula. (b) The effective diffusion coefficient $D^*$ (normalized
  by the medium diffusion coefficient $D$) as a function of $\epsilon$
  for $\lambda=50$. Circles - simulation results, solid curve - the LJ
  formula. (c) Same as (b) for $\lambda=2$. The dashed line depicts
  the function $(1-\epsilon)$.}
\label{fig:fig1}
\end{figure}

\section{Langevin dynamics simulations}

We simulate the dynamics of a particle of unity mass $m=1$ moving in a
system with friction coefficient $\alpha=1$ at constant temperature
$k_BT=1$. For this choice of parameters the ballistic time $\tau_b\sim
m/\alpha=1$ and ballistic length $l_b\sim\sqrt{mk_BT}/\alpha=1$. As a
numerical example, we consider the case where $\eta(x)=\cos(2\pi
x/\lambda)$ ($f(x)=-dU(x)/dx=k_BT\epsilon
\eta^{\prime}(x)/[1+\epsilon\eta(x)]$). The particle's trajectory
begins at $x=0$ with initial velocity which is drawn from the
equilibrium Maxwell-Boltzmann distribution, and is numerically
integrated using the algorithm of Gr{\o}nbech-Jensen and Farago
(G-JF)~\cite{gjf1,gjf2} with $dt=0.1$, which is an order of magnitude
smaller than $\tau_b$.  The numerical results presented here are based
on statistical averages of $2\times 10^8$ independent
trajectories. Fig.~\ref{fig:fig1}(a) shows the ratio between $\langle
x^2\rangle$ and $2t$ [see Eq.~(\ref{eq:effectived})] as a function of
$t$ for $\epsilon=0.5$ and $\lambda=50\gg l_b$. At asymptotically
large times, this ratio converges to the effective diffusion
coefficient $D^*\simeq 0.863$ which is indeed smaller than the medium
diffusion coefficient $D=k_BT/\alpha=1$. The open circles in
Fig.~\ref{fig:fig1}(b) show our computational results for $D^*/D$ as a
function of $\epsilon$, for $\lambda=50$. The results exhibit
excellent agreement with the solid line depicting the LJ formula
(\ref{eq:lifsonjackson2}) which, for the specific choice of $\eta(x)$
discussed here, gives $D^*/D=\sqrt{1-\epsilon^2}$. Very small
deviations from the LJ formula are observed for $\epsilon>0.6$ when
the potential barrier becomes larger [$\exp(-\beta \Delta U)\lesssim
  0.17$]. In contrast, the results for $\lambda=2$, which are plotted
in Fig.~\ref{fig:fig1}(c), exhibit agreement with LJ formula only for
$\epsilon<0.1$. This behavior is expected since LJ formula is derived
from the Smoluchowski equation, but the latter becomes invalid when
$\lambda$ is comparable to the ballistic length $l_b$. For
$\epsilon\rightarrow 1$, we observe that $D^*/D$ diminishes like
$(1-\epsilon)$, in accordance with Eq.~(\ref{eq:arrhen2}).

\begin{figure}[t]
\centering\includegraphics[width=0.45\textwidth]{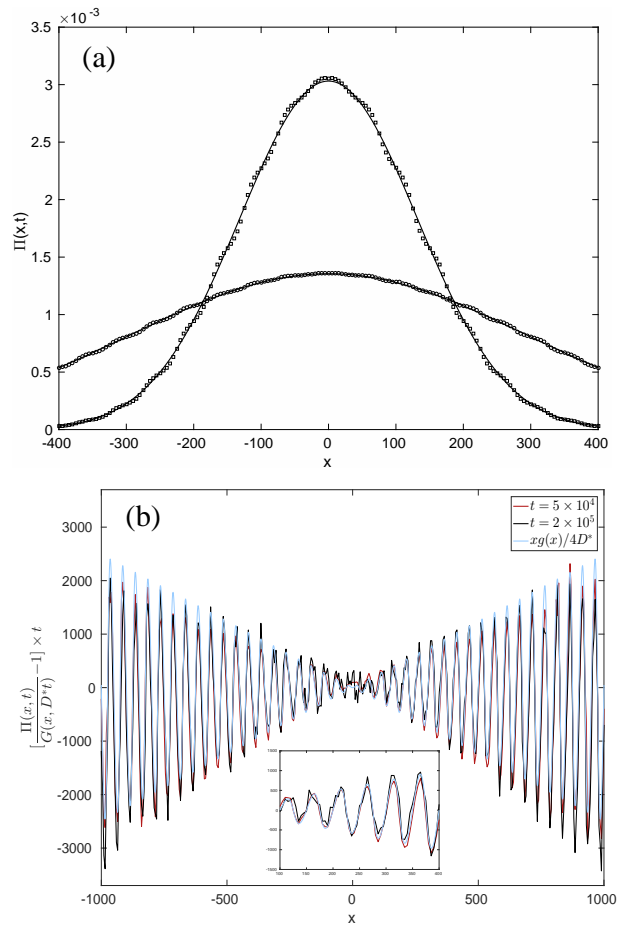}
\caption{(Color Online) (a) The function $\Pi(x,t)$ as a function of
  $x$ for $\epsilon=0.5$ and $\lambda=50$, at $t=10^4$ (squares) and
  and $t=5\times 10^4$ (circles). The solid curves depict the Gaussian
  form $G(x,D^*t)=\exp(-x^2/{4D^*t})/\sqrt{4\pi D^*t}$ with
  $D^*=0.863$. (b) The function $t\left[\Pi(x,t)/G(x,D^*t)-1\right]$
  as a function of $x$ for $t=5\times 10^4$ (red) and $t=2\times 10^5$
  (black). The light blue curve is the analytical approximation
  $xg(x)/(4D^*)$ with $g(x)$ given by Eq.~(\ref{eq:gtan}). The inset
  shows a magnification of the region with the best agreement between
  the analytical approximation and numerical results.}
\label{fig:fig2}
\end{figure}

Fig.~\ref{fig:fig2}(a) shows the PDF normalized by the Boltzmann
factor, $\Pi(x,t)=P(x,t)/[1+\epsilon\eta (x)]$, for $\epsilon=0.5$ and
$\lambda=50$ at $t=10^4$ (squares) and $t=5\times 10^4$ (circles). The
graphs indicate that at large times, the function $\Pi(x,t)$ is very
well approximated by a Gaussian form (solid lines)
$G(x,D^*t)=\exp(-x^2/{4D^*t})/\sqrt{4\pi D^*t}$, where $D^*=0.863$ is
the effective diffusion coefficient for the corresponding values of
$\epsilon$ and $\lambda$ (see above). This, however, is only an
approximation, and it is straightforward to check that
$P(x,t)=[1+\epsilon \eta(x)]G(x,D^*t)$ is {\em not}\/ a solution of
Eq.~(\ref{eq:smoluch2}). Careful inspection of Fig.~\ref{fig:fig2}(a)
reveals small undulations of $\Pi(x,t)$ with wavelength $\lambda$
around the Gaussian form. We thus speculate that the Gaussian form is
the leading term in an expression including additional terms, and
write $P(x,t)=[1+\epsilon
  \eta(x)]G(x,D^*t)\left[1+q_0(x)+q_1(x,t)\right]$, where $q_0(x)$ and
$q_1(x,t)$ denote, respectively, static and time-decaying corrections
exhibiting oscillations with periodicity $\lambda$. The static term
can be ruled out based on the simple argument that it contributes a
term equal to $G(x,D^*t)\,d\{[1+\epsilon\eta(x)]q_0^{\prime}(x)\}/dx$
on the right hand side of Eq.~(\ref{eq:smoluch2}), and this term
dominates the asymptotic behavior of $\partial_t P(x,t)$ at long
times. Since the Gaussian function $G(x,D^*t)$ is positive, it follows
that the long time limit of $\partial_t P(x,t)$ has the same sign as
the {\em static}\/ function
$d\{[1+\epsilon\eta(x)]q_0^{\prime}(x)\}/dx$ that oscillates between
positive and negative values \cite{footnote1}. This, however, is
impossible because the particle propagates to further distances with
time and the probability density cannot accumulate at any point in
space. Stated differently, for any $x_0$, the probability density
$P(x_0,t)$ must decay at large times, namely $\partial_t P(x_0,t)<0$
for some $t>t_0$. This property of $P(x,t)$ precludes the possibility
of a static correction $q_0(x)$ and allows only a time-decaying term
$q_1(x,t)$.

Insight into the form of the time-decaying correction can be gained
from Fig.~\ref{fig:fig2}(b) where we plot the function
$t\left[\Pi(x,t)/G(x,D^*t)-1\right]=tq_2(x,t)$ as a function of $x$
for $\epsilon=0.5$ and $\lambda=50$, at $t=5\times 10^4$ (red) and
$t=2\times 10^5$ (black). While the collapse of the data for the
different times in not perfect, it seems to indicate that the
time-decaying correction may have the form $q_1(x,t)\sim (x/t)g(x)$,
where $g(x)$ is a scaling function with periodicity $\lambda$. This
term decays at a rate $t^{1/2}$ faster than the leading Gaussian form
because of the scaling $x\sim(D^*t)^{1/2}$. (Larger values of $x$ need
not be considered because for $x^2\gg D^*t$ the PDF is practically
zero.) The solution can be further refined by introducing a series of
corrections
$P(x,t)=[1+\epsilon\eta(x)]G(x,D^*t)\left[1+q_1(x,t)+q_2(x,t)+\cdots\right]$,
with each term decaying $t^{1/2}$ faster than the previous one. Here,
we attempt to find the first two leading corrections and, therefore,
consider the following form
\begin{widetext}
\begin{equation}
  P(x,t)=\left\{\left[1+\epsilon
    \eta(x)\right]G(x,D^*t)\left[1+\frac{\lambda xg(x)}{4D^*
      t}Q_1\left(\frac{x^2}{4D^* t}\right)+\frac{\lambda^2
      h(x)}{4D^*t}Q_2\left(\frac{x^2}{4D^* t}\right)\right]\right\},
  \label{eq:solution}
\end{equation}
\end{widetext}
where $g(x)$ and $h(x)$ are two dimensionless functions with
periodicity $\lambda$, while $Q_1$ and $Q_2$ are polynomials in
$x^2/4D^* t$.  This solution constitutes the {\em asymptotic}\/
solution to Smoluchowski equation (\ref{eq:smoluch2}) at large times
$t\gg \tau_b$, up to order $G(x,D^*t)/t\sim 1/t^{3/2}$.

\section{The probability distribution}

\subsection{The leading asymptotic correction}

We consider a symmetric system with an even function $\eta(x)$, which
implies that the PDF is symmetric and, therefore, $g(x)$ must be an
odd function while $h(x)$ is even. Both scaling functions are periodic
with periodicity $\lambda$ and have a finite amplitude.  Thus,
$\eta^{\prime}(x)$, as well as $g^{\prime}(x)$ and $h^{\prime}(x)$,
are all of order $1/\lambda$. With this in mind, we substitute the
solution (\ref{eq:solution}) into Eq.~(\ref{eq:smoluch2}). On the
right hand side, we find terms that scale as like $G(x,D^*t)(x/\lambda
t)$. These terms must cancel each other, which occurs provided that
(i) $Q_1={\rm Const}$ (which, by a proper definition of $g(x)$, can be
arbitrarily set to unity), and (ii) the scaling function $g(x)$
satisfies the ordinary differential equation
\begin{equation}
  \left[1+\epsilon\eta(x)\right]\lambda g^{\prime\prime}(x)
  +\epsilon\eta^{\prime}(x)\left[\lambda g^{\prime}(x)-2\right]=0.
\end{equation}
Taking into account that $g(x)$ is an odd function and therefore
$g(0)=0$, we readily arrive at the solution
\begin{equation}
  \lambda g(x)=kI\left[\frac{1}{1+\epsilon\eta(x)}\right]+2x,
  \label{eq:gfunction}
\end{equation}
where $I[y(x)]$ denotes the primitive function of $y(x)$ with
$I(x=0)=0$. The constant $k$ can be found from the requirement that
$g(x)$ is periodic. Thus,
$\int_0^{\lambda}g^{\prime}(x)dx=g(\lambda)-g(0)=0$, which gives
\begin{equation}
  k=-\frac{2}{\left\langle
    \left[1+\epsilon\eta(x)\right]^{-1}\right\rangle}.
  \label{eq:constantk}
 \end{equation}

\subsection{The next asymptotic correction}

 We now proceed and find the function $h(x)$ and the polynomial $Q_2$
 by comparing terms of the form $G(x,D^*t)(x^2/4D^*t)^n/t$ (where $n$
 is an integer), on both sides of Eq.~(\ref{eq:smoluch2}). From this
 comparison, we readily conclude that $Q_2\left(x^2/4D^*
 t\right)=[1+bx^2/(4D^*t)]$. This leaves us with two different
 differential equations for the scaling function $h(x)$ which cannot
 be solved simultaneously unless we set $b=-2$, in which case the
 equations coincide and read
\begin{eqnarray}
  &(1+\phi)&\!\!\![1+\epsilon\eta(x)]\left[\lambda^2
    h^{\prime\prime}(x) +2\lambda
    g^{\prime}(x)\right]+\\ &(1+\phi)&
  \!\!\!\epsilon\eta^{\prime}(x)\left[\lambda^2
    h^{\prime}(x)+\lambda g(x)\right]
  =2\phi(\left[1+\epsilon\eta(x)\right],\nonumber
\end{eqnarray}
where $\phi=D/D^*-1$. By defining the function
$\tilde{h}(x)=\left[1+\epsilon\eta(x)\right]\left[\lambda^2
  h^{\prime}(x)+\lambda g(x)\right]$, we arrive at the simple equation
\begin{equation}
  \tilde{h}^{\prime}(x)=-\left[\frac{2}{1+\phi}+k\right]
  -\frac{2\epsilon\eta(x)}{1+\phi},
  \label{eq:hode}
\end{equation}
which can be integrated once to yield $\tilde{h}(x)$, from which the
scaling function $h(x)$ can be derived by performing yet another
integration in $x$. Importantly, the fact that $h(x)$ is an even
periodic function with a finite amplitude imposes the relationship
$2/(1+\phi)+k=0$ from which $D^*$ can be deduced. Using
Eq.~(\ref{eq:constantk}) we find
\begin{equation}
  D^*=\frac{D}{1+\phi}=-\frac{kD}{2}=\frac{D}{ \left\langle
    \left[1+\epsilon\eta(x)\right]^{-1}\right\rangle},
\end{equation}
which is identical to the LJ formula (\ref{eq:lifsonjackson2}). The
scaling function $h(x)$ is given by
\begin{equation}
  \lambda^2 h(x)=kI\left\{\frac{I\left[\epsilon
      \eta(x)\right]}{1+\epsilon\eta(x)}-
  I\left[\frac{1}{1+\epsilon\eta(k)}\right]\right\}-x^2+\lambda^2 C,
  \label{eq:hfunction}
\end{equation}
where the constant $C$ is determined by the normalization condition
$\int_{-\infty}^{\infty} P(x,t)dx=1$ (which must be satisfied up to an
order of $1/t$).

\subsection{Comparison with simulation results} 

Returning to our simulation results above for $\eta(x)=\cos(2\pi
x/\lambda)$, the corresponding scaling function $g(x)$ can be found
\begin{equation}
  g(x)=-\frac{2}{\pi}\arctan
  \left[\sqrt{\frac{1-\epsilon}{1+\epsilon}}
    \tan\left(\frac{\pi x}{\lambda}\right)\right]+\frac{2x}{\lambda},
  \label{eq:gtan}
\end{equation}
where the $\arctan$ function is interpreted such that it returns a
value between $-\pi/2$ and $\pi/2$ which is then shifted by an integer
number of $\pi$ in order that $g(\lambda/2+n\lambda)=0$ (where $n$ is
an integer). We were unable to analytically perform the integration in
Eq.~(\ref{eq:hfunction}), necessary for finding a closed-form
expression for the scaling function $h(x)$; however, we take advantage
of the fact that it represents a correction to the Gaussian form of
$\Pi(x,t)$ which is asymptotically smaller than the one involving the
function $g(x)$ [see Eq.~(\ref{eq:solution})], and use the
approximation $t\left[\Pi(x,t)/G(x,D^*t)-1\right]\simeq
xg(x)/(4D^*)$. This approximation, which is plotted in
Fig.~\ref{fig:fig2}(b) in light blue, shows a good fit to the
simulation results. The deviations, which can be attributed to the
higher order correction term in Eq.~(\ref{eq:solution}), are
particularly small for $x^2\simeq 2D^*t$ - see inset in
Fig.~\ref{fig:fig2}(b).

\section{Summary and future outlook} 

In summary, we derived an asymptotic (at large times) expression for
the PDF of a particle diffusing in a periodic potential energy
landscape $U(x)$. The solution, which is given by
Eq.~(\ref{eq:solution}), is correct to order $1/t^{3/2}$. Faster
decaying corrections can, in principle, be systematically derived by
comparing, on both sides of Eq.~(\ref{eq:smoluch2}), terms of order
$1/t^2$, $1/t^{5/2}$~$\ldots\,$. This will require solving
increasingly complicated differential equations involving the scaling
functions of slower decaying terms.

We conclude by noting that our approach to solving the Smoluchowski
equation can be used, with only simple modifications, to solving other
closely-related diffusion equation. These include, for instance, the
equation describing Brownian motion in a tilted periodic potential,
i.e., when a particle diffuses under the action of both a periodic
potential $U(x)$ and a constant force $f$. In this case, the
Smoluchowski equation reads
\begin{align}
  &\frac{\partial P(x,t)}{\partial t}=D\frac{\partial}{\partial
    x}\left\{e^{-\beta (U(x)-fx)}\frac{\partial}{\partial x}
  \left[e^{\beta
      (U(x)-fx)}P(x,t)\right]\right\}\nonumber\\
  &=-\frac{f}{\alpha}\frac{\partial
    P}{\partial x}+D\frac{\partial}{\partial
    x}\left\{\left[1+\epsilon\eta(x)\right]\frac{\partial}{\partial x}
  \frac{P(x,t)}{\left[1+\epsilon\eta(x)\right]}\right\},
  \label{eq:smoluchforce}
\end{align}
where $\alpha=k_BT/D$, while $\eta(x)$ and $\epsilon$ are defined
similarly to Eqs.~(\ref{eq:eta}) and (\ref{eq:epsilon}).  The two main
differences with respect to the case $f=0$ considered previously: (i)
The Gaussian ``moves'' at constant velocity $f/\alpha^*$, where
$\alpha^*\neq\alpha$ is the effective friction coefficient satisfying
$\alpha^*\rightarrow k_BT/D^*$ for $f\rightarrow 0$
\cite{gunther79}. (ii) Corrections to the Gaussian form also include a
static term $q_0(x)$ missing when $f=0$, with oscillations having the
wavelength $\lambda$ of the potential $U(x)$. Thus, we speculate that
the solution takes the form
\begin{align}
P(x,t)&=\left[1+\epsilon
  \eta\left(x\right)\right]G\left(x-\frac{f}{\alpha^*}t,D^*t\right)\times
\nonumber \\ &\times \left[1+q_0(x)+q_1(x,t)+q_2(x,t)+\cdots\right],
\label{eq:solutionforce}
\end{align}
where $G=\exp[-(x-tf/\alpha^*)^2/4D^*t]/\sqrt{4\pi D^*t}$ denotes the
``running'' Gaussian, while $q_1$ and $q_2$ are the time decaying
corrections of order $1/t^{1/2}$ and $1/t$, respectively, having the
same general form as in Eq.~(\ref{eq:solution}) of the main text.

Upon substituting the solution (\ref{eq:solutionforce}) into
Eq.~(\ref{eq:smoluchforce}) and comparing terms of similar order, we
first arrive at the following differential equation for the static
term $q_0(x)$
\begin{equation}
  \frac{\partial}{\partial
    x}\left\{\left[1+\epsilon\eta\left(x\right)\right] \left[\beta
    f+\beta
    fq_0\left(x\right)-q_0^{\prime}\left(x\right)\right]\right\}=0.
\end{equation}
From the requirements that (i) $q_0(x)$ does not diverge exponentially
for $x\rightarrow\infty$, and that (ii) $P(x,t)$ is normalized to
unity at any time (including for $t\rightarrow\infty$, when the
time-decaying terms become irrelevant), we arrive at
\begin{equation}
q_0(x)=\beta fe^{\beta fx}I\left[e^{-\beta
    fx}\left(\frac{c_1}{1+\epsilon\eta(x)}+1\right)\right],
\label{eq:q0}
\end{equation}
where (as above) $I$ denote a primitive function, and
\begin{equation}
  c_1=\left\{\beta
  f\left\langle\left[1+\epsilon\eta\left(x\right)\right]e^{\beta
    fx}I\left[\frac{e^{-\beta fx}}{1+\epsilon
      \eta(x)}\right]\right\rangle\right\}^{-1}.
\label{eq:c1}
\end{equation}

The first time-decaying correction $q_1(x)$ can be now found by
solving the equation obtained from the terms that scale as
$1/t^{1/2}$. This equation involves the already found static scaling
function $q_0(x)$. Then, the second correction, $q_2(x)$, can be found
from the equation corresponding to the terms proportional to $1/t$,
which involves $q_0(x)$ and $q_1(x)$ (and so on). Notice that the form
of $q_0$ (\ref{eq:q0}) does not give any information on the effective
parameters $\alpha^*$ and $D^*$. These will be found from the
equations for $q_1$ and $q_2$, respectively.

Acknowledgments: This work was supported by the Israel Science
Foundation (ISF) through grant No.~991/17.


\begin{thebibliography}{99}
\bibitem{footnote1} The last statement is incorrect when
  $q_0^{\prime}(x)=[1+\epsilon\eta(x)]^{-1}$. However, this
  possibility can be ruled out since in this case the average slope of
  $q_0$ does not vanish, which implies that $q_0(x)$ and, therefore,
  also the probability density $P(x,t)$ obtain large negative values
  for some $x$.
\end{thebibliography}
\end{document}